\documentclass[cpp,a4paper,fleqn%
]{w-art}
\usepackage{times,cite,w-thm}
\theoremstyle{plain}

\theoremstyle{definition}

\usepackage[]{graphicx}
\usepackage{amssymb}
\usepackage{amsfonts}

\renewcommand{\vec}[1]{\mathbf{#1}}

\begin{document}
\DOIsuffix{theDOIsuffix}
\Volume{46}
\Month{01}
\Year{2007}
\pagespan{1}{}
\Receiveddate{XXXX}
\Reviseddate{XXXX}
\Accepteddate{XXXX}
\Dateposted{XXXX}



\title[Ion Potential in Non-Ideal Dense Quantum Plasmas]{Ion Potential in Non-ideal Dense Quantum Plasmas}


\author[Zh.A. Moldabekov et al.,]{Zh.A.~Moldabekov$^{1,2}$}
\author[]{S.~ Groth$^{1}$}
\author[]{T.~Dornheim$^{1}$}
\author[]{M.~Bonitz$^{1}$}
\author[]{T.S.~Ramazanov$^{2}$}

\address[]{$^1$Institut f{\"u}r Theoretische Physik und Astrophysik, Christian-Albrechts-Universit\"at zu Kiel,
D-24098 Kiel, Germany}
\address[]{$^2$Institute for Experimental and Theoretical Physics, Al-Farabi Kazakh National University,\\ 71 Al-Farabi av.,  
 050040 Almaty, Kazakhstan}

\begin{abstract}
The screened ion potential in non-ideal dense quantum plasmas is investigated by invoking the Singwi-Tosi-Land-Sj\"olander approximation for the electronic local field correction at densities $r_s\lesssim 2$ and degeneracy parameters  $\theta\lesssim 1$, where $r_s$ is the ratio of the mean inter-particle distance to the first Bohr radius, and $\theta$ is the ratio of the thermal energy to the Fermi energy of the electrons.
Various cross-checks with ion potentials obtained from ground state quantum Monte-Carlo data, the random phase approximation, as well as with existing analytical models are presented.
Further, the importance of the electronic correlation effects for the dynamics in strongly coupled ionic subsystems for $0.1\leq r_s\leq 2$ is discussed.

\end{abstract} 
\maketitle
\section{Introduction}

Even though many advances regarding the modeling and simulation of non-equilibrium electron-ion plasmas could be achieved recently~\cite{Militzer, Filinov, Lambert, Holst}, a fully self-consistent treatment has yet been prevented by the fact that electronic quantum and
spin effects together with strong ionic correlations must be taken into account simultaneously. 
The main difficulties arises from the vastly different time scales of the electrons and ions, a direct consequence of their different masses. 
A possible way to overcome this obstacle is given by the multi-scale approach proposed by Ludwig et al.~\cite{Patrick},
which relies on a linear response treatment of the electrons that is justified in the case of weak electron-ion coupling. 
The key of this multi-scale approach is that the fast electron kinetics are incorporated into an effective screened potential of the heavy ions, where the screening is provided by the electrons via a proper dielectric function.
Recently, we have performed the analysis of the screened ion potential in the random phase approximation (RPA)~\cite{POP2015}, which is naturally expected to be valid in the weak coupling regime. 
In this work, we extend these investigations to the case of non-ideal quantum electrons by making use of local field corrections that are computed within the Singwi-Tosi-Land-Sj\"olander formalism (STLS)~\cite{Tanaka} and 
from quantum Monte Carlo simulations~\cite{Corradini}. 

The properties of the electrons in a dense partially (or fully) degenerate plasma are commonly characterized by the density parameter $r_s=a/a_B$ and the quantum degeneracy parameter $\theta=k_BT_e/E_F$, where 
$a=(3/(4\pi n))^{1/3}$, 
with $a_B$ being the Bohr radius, $T_e$ the temperature of the electrons, $k_B$ the Boltzmann constant, and $E_F$ the Fermi energy of the electrons.

In the past, Bennadji et al.~\cite{Bennadji} calculated the screened potential in the STLS approximation at fixed temperature $T=10^4~{\rm K}$ and densities ranging from $10^{19}$ to $10^{26}~{\rm cm^{-3}}$. Nevertheless, a thorough investigation of the screened ion potential in dependence of the ion distance as well as a careful test of the accuracy of the STLS approximation have not been performed. Moreover, in Refs.~\cite{Fletcher, Ma}, the screened ion-ion interaction potential was described by a model consisting of a ``Yukawa + a short-range repulsive potential", which has been designed to fit the results of warm dense matter studies within the density functional theory formalism.
However, this model was later criticized by Harbour et al.~\cite{Harbour}, who investigated the compressibility, phonons, and  electrical conductivity of warm dense matter on the basis of a neutral-pseudoatom model. The general approach of utilizing a screened pair interaction potential has been considered and implemented in numerous previous works~\cite{Deutsch, Ramazanov1, Ramazanov2} to study various physical properties of semiclassical weakly coupled plasmas. Recently, Baalrud and Daligault could extend the applicability of an effective pair interaction potential to the description of strongly coupled classical plasmas~\cite{Baalrud}.  
 
Here, we compare different models for the screened ion potential in quantum plasmas, for $\theta\lesssim 1$, and focus on the influence of electronic correlations effects.
Note that we do not take into account electron-ion recombination effects (possible bound states). 
This is justified by first-principles path-integral Monte Carlo simulations~\cite{Bonitz} which indicate that, due to the so-called Mott effect or pressure ionization, bound states break up in the density range around $r_s=1.5...2$. 
Therefore, we only consider density parameters $r_s\lesssim2$.
These high densities are in particular relevant for non-ideal (moderately) degenerate plasmas found in inertial confinement fusion experiments~\cite{Hurricane, Cuneo, Gomez, Hoffmann1}, and neutron star atmospheres~\cite{Potekhin}. 

This paper is structured as follows: in Sec.~\ref{sec:2} the theoretical formalism is presented, and the results for the screened ion potentials are shown in Sec.~\ref{sec:3}. In the last section we summarize our findings and give a short conclusion. 

\section{Screened ion potential}\label{sec:2}
 The screened  potential $\Phi$ of an ion with charge $Q=Z|e|$ can be calculated using the well-known formula~\cite{Patrick}:
\begin{equation} \label{POT_stat}
\Phi(\vec r)   = \int\!\frac{\mathrm{d}^3k}{2 \pi^2 } \frac{Q^2}{k^2 \epsilon(\vec k, \omega=0)} \,\, e^{i \vec k \cdot \vec r} \quad,
\end{equation}
with $\epsilon(\vec{k},\omega)$ being the dielectric response function of the electrons 
\begin{equation}\label{dynamic_epsilon}
 \epsilon(\mathbf{k},\omega) = 1 - \frac{\chi_0(\mathbf{k},\omega)}{k^2/(4\pi e^2)+G(\mathbf{k,\omega})\chi_0(\mathbf{k},\omega)}.
\end{equation}
According to Eq.~(\ref{POT_stat}), the screening effects of the electrons are entirely determined by the static limit of the dielectric function
\begin{equation}\label{static_epsilon}
 \epsilon(\mathbf{k},\omega=0) = 1 - \frac{\chi_0(\mathbf{k},0)}{k^2/(4\pi e^2)+G(\mathbf{k},0)\chi_0(\mathbf{k},0)}.
\end{equation}
Here, $\chi_0$ denotes the finite temperature ideal density response function of the electron gas~\cite{quantum_theory}. Further, all correlation effects are incorporated in the so-called local field correction $G$, so that the dielectric function in the RPA is given by setting $G=0$ in Eq.~(\ref{static_epsilon})
\begin{equation}
\epsilon_\text{RPA}(\mathbf{k},0) = 1-\frac{4\pi e^2}{k^2}\chi_0(\mathbf{k},0).
\end{equation}
A highly successful way to approximately determine the static local field correction and, thereby, going beyond the RPA is provided by the self-consistent static STLS scheme~\cite{stls}, which is based on the ansatz
\begin{equation}
G(\mathbf{k},\omega)\approx G^\textnormal{STLS}(\mathbf{k},0) = -\frac{1}{n} \int\frac{\textnormal{d}\mathbf{k}^\prime}{(2\pi)^3}
\frac{\mathbf{k}\cdot\mathbf{k}^\prime}{k^{\prime 2}} [S^{\text{STLS}}(\mathbf{k}-\mathbf{k}^\prime)-1]\;,
\label{G_stls}
\end{equation}
where the static structure factor $S^{\textnormal{STLS}}$ can be computed according to the fluctuation-dissipation theorem as
\begin{equation}\label{flucDis}
 S^\text{STLS}(\mathbf{k}) = -\frac{1}{\beta n}\sum_{l=-\infty}^{\infty} \frac{k^2}{4\pi e^2}\left(\frac{1}{\epsilon(\mathbf{k},z_l)}-1\right)\ .
\end{equation}
Note that the integration over the frequencies in the fluctuation-dissipation theorem has been re-casted in a summation over the Matsubara frequencies $z_l=2\pi il/\beta\hbar$. Following~\cite{stls}, the complex valued dielectric function is straightforwardly evaluated via Eq.~(\ref{static_epsilon}) from the corresponding finite temperature ideal response function and $G^\text{STLS}$. Thus, Eqs.~(\ref{dynamic_epsilon}), (\ref{G_stls}), and (\ref{flucDis}) form a closed set of equations that can be solved iteratively until self-consistency is reached, which finally yields the converged dielectric funciton $\epsilon^\text{STLS}(\mathbf{k},0)$. For the RPA,  STLS and quantum Monte Carlo result of the static dielectric function we readily obtain the corresponding screened potentials via Eq.~(\ref{POT_stat}). 

A widely used analytical model that incorporates screening effects is given by the Yukawa type potential:
\begin{equation}\label{Yukawa}
\Phi_{Y}(r;n,T)=\frac{Q}{r}\, e^{-k_{Y}r},
\end{equation}
with $k_{Y}^2(n,T)=\frac{1}{2} k_{TF}^2 \theta ^{1/2} I_{-1/2}(\beta \mu)$,
$k_{TF}=\sqrt{3}\omega_{p}/v_{F}$ (the Thomas-Fermi wave number), and $I_{-1/2}$ being the Fermi integral of order $-1/2$ (see below). 
In the context of quantum plasmas, Eq.~(\ref{Yukawa}) is often referred to as the Thomas-Fermi potential (TF).
The inverse screening length $k_Y$ interpolates between the Debye and Thomas-Fermi expressions in the non-degenerate and zero temperature limits, respectively.

Another analytical model has been provided by Stanton and Murillo (SM)~\cite{Murillo}:
\begin{align}\label{eq:sm>}
\phi^{SM>}(r;n,T) = \frac{Q}{r}\left[\cos(k'_- \cdot r)+ b' \sin(k'_- \cdot r )\right] \, e^{-k'_+ \cdot r},
\end{align}
where $k'_\pm = k_{TF}(\sqrt{\alpha^{SM}} \pm 1)^{1/2} / \sqrt{\alpha^{SM}}$, $\alpha^{SM} = 3 \sqrt{8 \beta} \lambda {I'}\!\!_{-1/2}(\eta_0) / \pi$, $\lambda = 1/9$, $b' = 1/\sqrt{\alpha^{SM} - 1}$. Here,  $I_p(\eta) = \int_0^\infty dx\; x^p / (1 + e^{x-\eta})$ denotes the Fermi integral and $I'_p(\eta)$ its derivative with respect to $\eta$. In addition, $\eta_0$ is determined by the normalization, $n_0 = \sqrt{2} I_{1/2}(\eta_0) / \pi^2\beta^{3/2}$ with the inverse temperature $\beta$, and the inverse Thomas-Fermi screening length at finite temperatures is given by $k_{TF} = (4I_{-1/2}(\eta_0)/\pi\sqrt{2\beta})^{1/2}$.

For the case $\alpha^{SM} < 1$ the SM potential takes a different form:
\begin{align}\label{eq:sm<}
\phi^{SM<}(r;n,T) = \frac{Q}{2r}\left[(1+b)\, e^{-k_+ r} + (1-b)\, e^{-k_- r}\right],
\end{align}
where $b = 1/\sqrt{1 - \alpha^{SM}}$ and $k_\pm = k_{TF}(1\mp \sqrt{1-\alpha^{SM}})^{1/2} / \sqrt{\alpha^{SM}/2}$. We note that for the ground state ($\theta=0$) the screened potentials~(\ref{eq:sm>}) and (\ref{eq:sm<}) were derived by Akbari-Moghanjoughi \cite{Akbari}.

The SM potential was derived starting from the Thomas-Fermi model with the first order gradient correction to the non-interacting free energy density functional \cite{Murillo}. In Ref.~\cite{POP2015}, it was shown that this corresponds to the second order result of the long wavelength expansion of the inverse ideal response function. Whereas the TF potential (\ref{Yukawa}) can be derived considering  the lowest order result, which is given by neglecting all $k$-dependent terms in the long wavelength expansion of the inverse response function in the RPA. Therefore, both TF and SM potentials correctly describe the screening of the ion potential at large distances but completely neglect non-ideality (correlation) effects.

\section{Results}\label{sec:3}
\begin{figure}[t]
\centering
\includegraphics[width=0.7\textwidth]{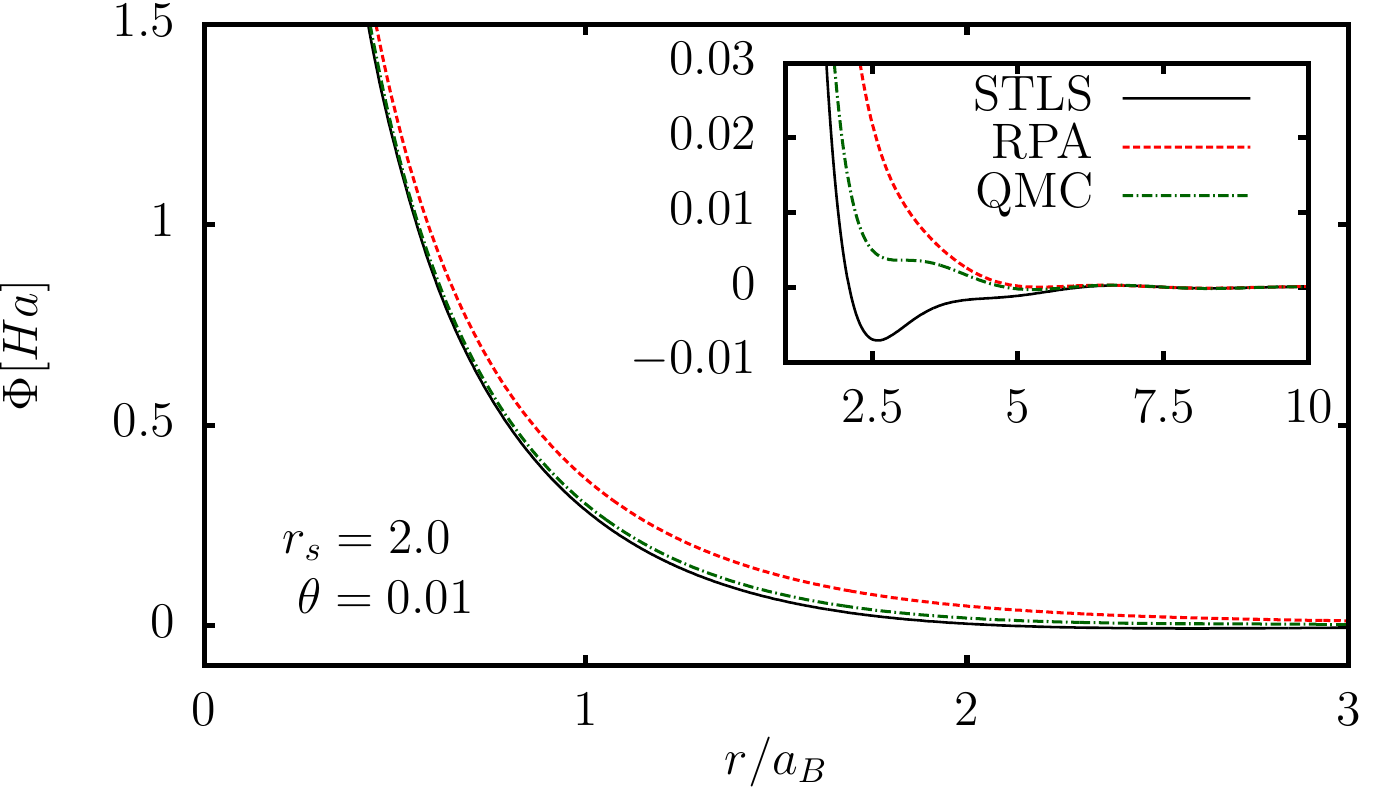}\hspace{2pc}%
\caption{\label{fig:1} STLS, RPA, and QMC \cite{Corradini} results for the screened ion potential at $r_s=2.0$ and $\theta=0.01$.}
\end{figure}

\begin{figure}[h]
 \centering
  \includegraphics[width=0.9\textwidth]{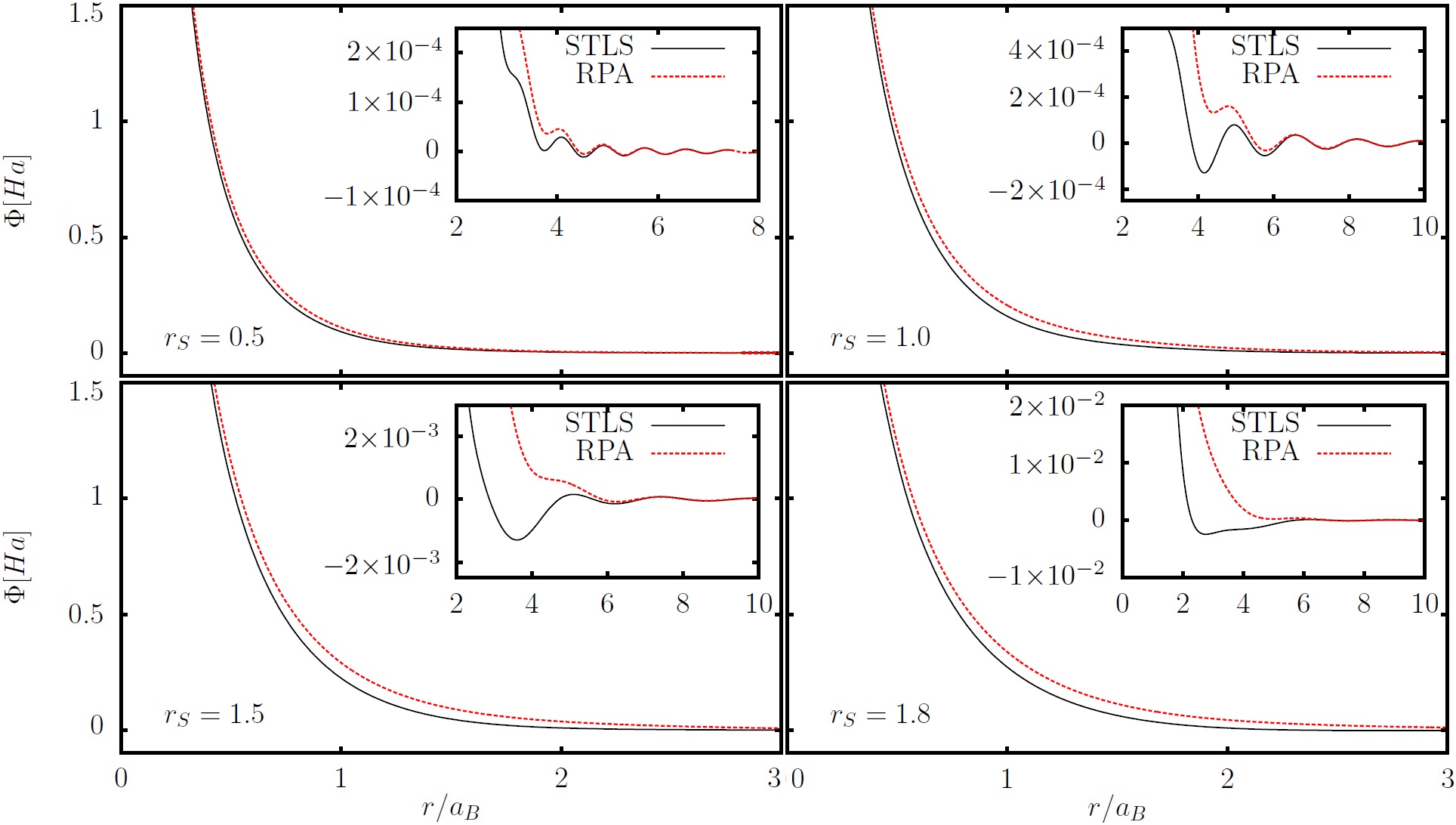}
  \caption{STLS and RPA results for the screened ion potential for $\theta=0.01$ at different values of the density parameter. 
The impact of the electronic correlations on the charge screening becomes stronger with decreasing density. 
 The negative minimum in the STLS ion potential can be distinguished from the Friedel oscillations only at $r_s>1$. }
 \label{fig:2}
 \end{figure}

First, in Fig.~\ref{fig:1} we determine the accuracy regarding the treatment of correlation effects of the STLS approximation by comparing it to the parametrization of the exact ground state quantum Monte Carlo (QMC) data~\cite{Corradini} at $r_s=2$. Clearly, one can see that electronic correlations lead to a stronger screening of the test charge potential in comparison to the RPA result.
Tested against the exact QMC data,  the STLS approximation provides a qualitatively good description of the correlation effects in the screened potential. In particular, the behavior at short distances, $r\lesssim a$, is described much better than by the RPA ansatz.   However, unlike the ion potential based on the QMC data, which is monotonically decreasing, the STLS ion potential exhibits a shallow negative minimum ($\Phi_{\rm min}<10^{-2}~{\rm Ha}$) at $r\simeq 2.5~a_B>a$.
This appears to be an artifact of the STLS approximation.
In Fig.~\ref{fig:2} the STLS and RPA potential are shown for different values of the density parameter $r_s$ at fixed $\theta=0.01$ (very close to the ground state).
With increasing density the negative minimum in the ion potential becomes less pronounced, i.e the absolute value of $\Phi_{\rm min}$ decreases, and finally becomes indistinguishable from the so-called Friedel oscillations at $r_s\leq 1$. As expected, the difference between the STLS approximation and the RPA reduces with increasing density. 

\begin{figure}[t]
 \centering
  \includegraphics[width=0.48\textwidth]{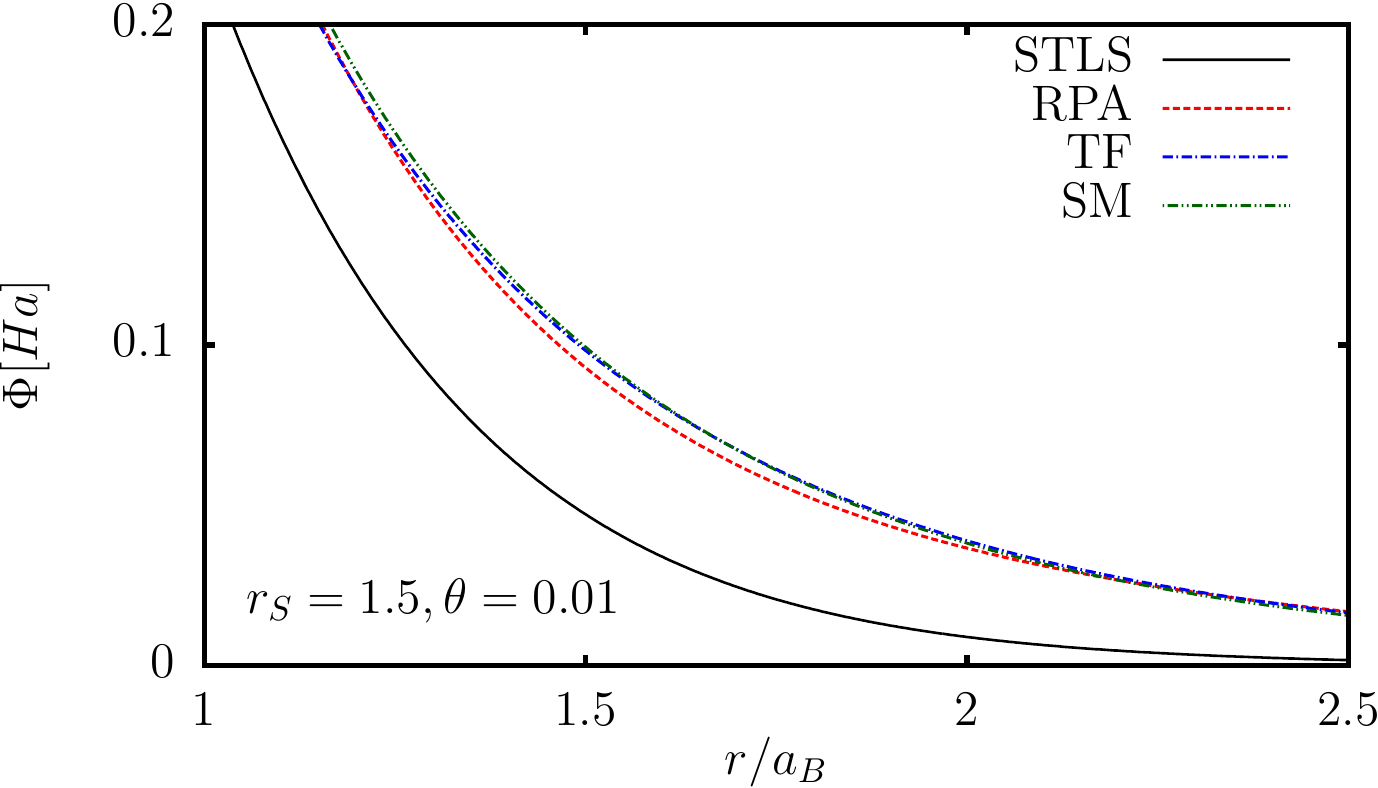}~\hspace{-2mm}\begin{scriptsize}a)\end{scriptsize}
  \includegraphics[width=0.48\textwidth]{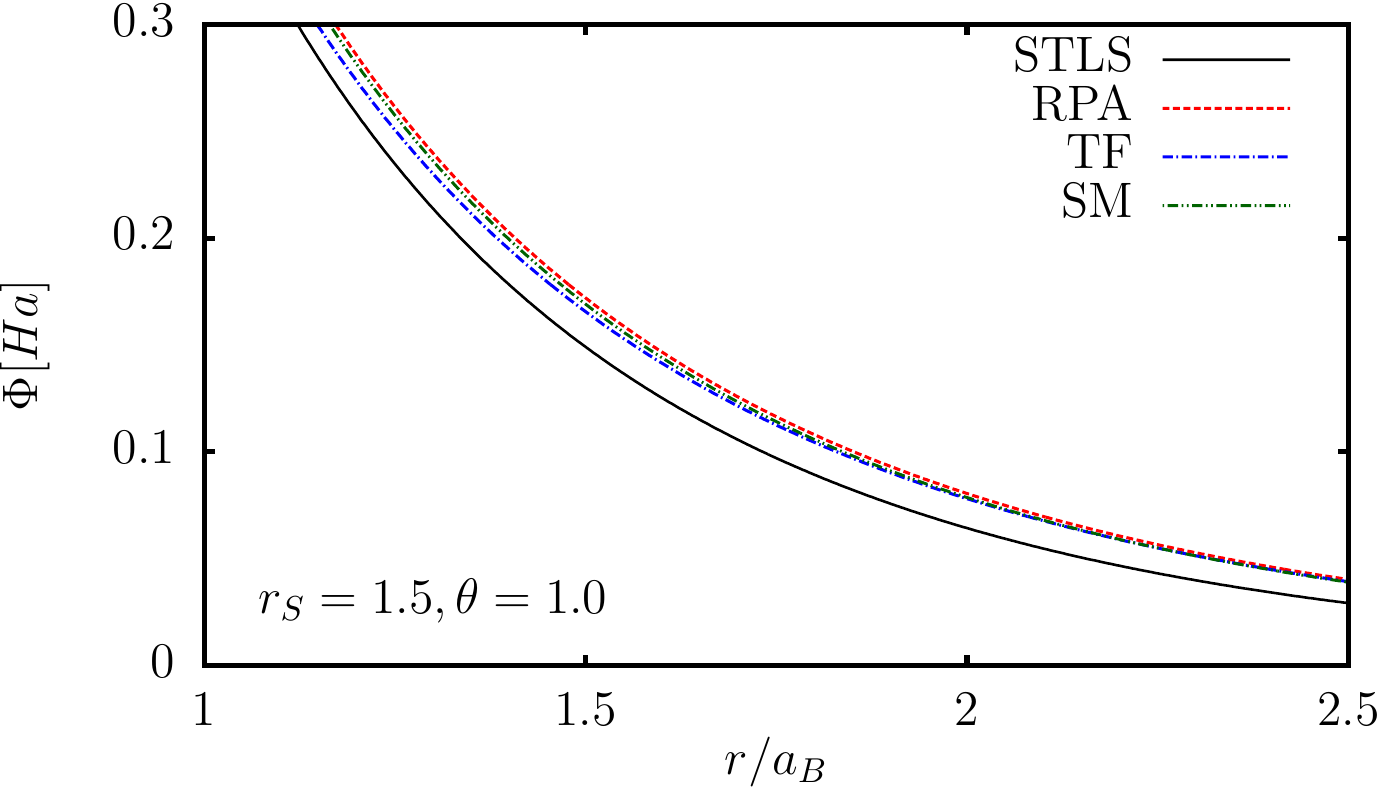}~\hspace{-2mm}\begin{scriptsize}b)\end{scriptsize}
  \caption{STLS and RPA results in comparison with the TF (\ref{Yukawa}) and SM (\ref{eq:sm>}), (\ref{eq:sm<}) potentials at a) $\theta=0.01$ and b) $\theta=1.0$ for $r_s=1.5$.}
 \label{fig:4}
 \end{figure}

In Fig.~\ref{fig:3} the comparison of the ion potentials in the STLS approximation and RPA are presented for different values of the degeneracy parameter $\theta$ at $r_s=1.5$. 
Since the effect of correlations becomes less important with increasing temperature, the difference between the STLS and RPA reduces with temperature.
In Fig.~\ref{fig:4} a) and b), we also plot the TF (\ref{Yukawa}) and SM  (\ref{eq:sm<}) potentials. 
While TF, SM and RPA can barely be distinguished, they all differ significantly from the STLS potential since only here correlation effects are taken into account.

\begin{figure}[h]
 \centering
 \includegraphics[width=0.85\textwidth]{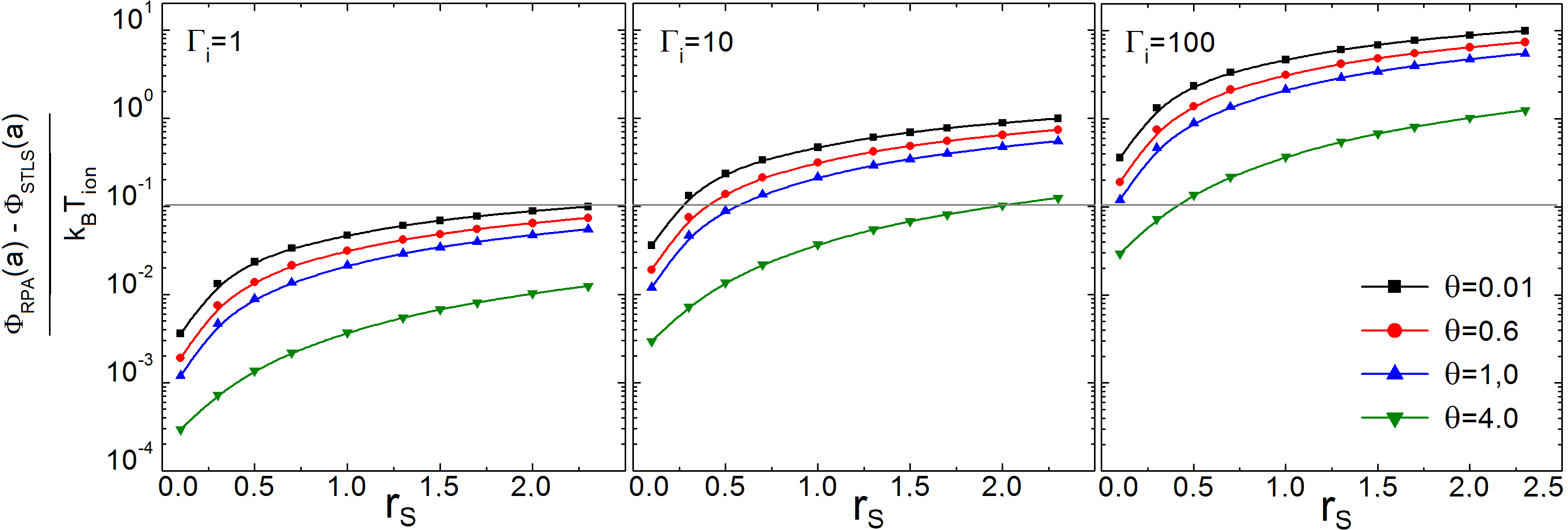}\hspace{2pc}%
 \caption{ Difference between RPA and  STLS results for the ion potential at the mean interparticle distance in units of the thermal energy of the ion, for different values of the ion coupling parameter. The effect of the electrons non-ideality on the ion dynamics can be significant if the ionic subsystem is 
 strongly correlated, i.e. $\Gamma_i>1$.}
 \label{fig:5}
 \end{figure}

Finally, to assess the relative importance of electronic correlations in the screened ion potential, in Fig.~\ref{fig:5}, we show the ratio of the difference between the STLS and RPA potential at the mean inter particle distance ($a=r_s\times a_B$) to the characteristic thermal energy of the ions, $k_BT_{\rm ion}=Z^{5/3}/(r_s \Gamma_i)~[{\rm Ha}]$, for different values of the degeneracy and ion coupling parameter
 $\Gamma_i=Q^2/a_ik_BT_{\rm ion}$, where $a_i=(3Z/(4\pi n))^{1/3}$ is the mean distance between ions. The ion coupling parameter can be written as $\Gamma_i=2\left(\frac{4}{9\pi}\right)^{2/3} \frac{r_s}{\theta} \frac{T_e}{T_{\rm ion}}$. The considered values of $\Gamma_i$ are 1, 10 and 100, where we take $Z=1$. At $\Gamma_i\geq 10$, the difference in the ion potential due to electronic correlations is comparable to the thermal energy of the ions even at $r_s<1$.
Here we assumed that $T_e\neq T_{\rm ion}$, which is reasonable for many warm dense matter experiments. For $Z=1$,   the ratio of the electron temperature to the temperature of ions can be expressed as $T_e/T_{\rm ion}\simeq 1.84 \times (\theta/r_s)\, \Gamma_i$, which for $\theta=0.6$, $r_s=1$, and $\Gamma_i\leq100$ corresponds to the values $T_e/T_{\rm ion} \lesssim 110$.     
 
\hspace{20mm}
\begin{figure}[t]
\centering
\includegraphics[width=0.5\textwidth]{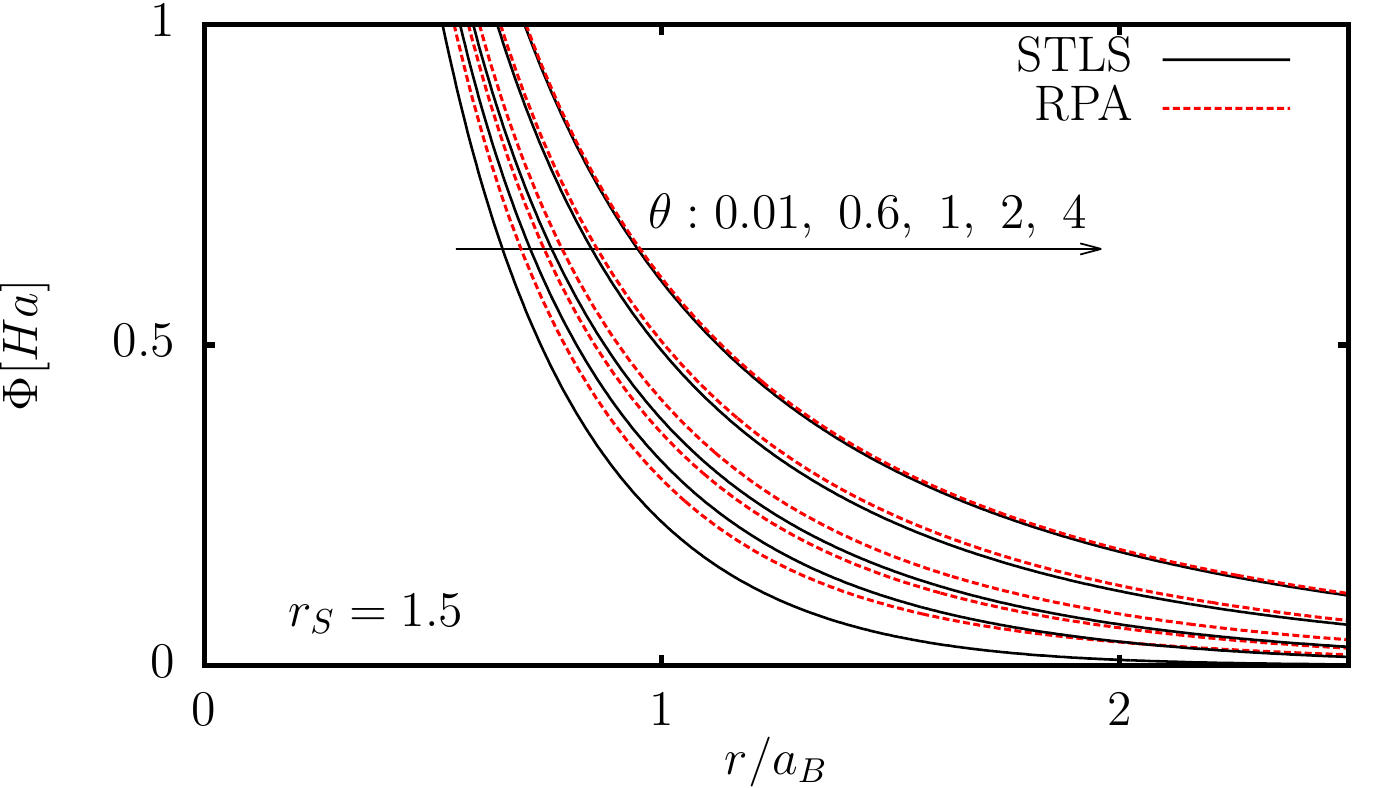}\hspace{2pc}%
\begin{minipage}[b]{16pc}\caption{\label{fig:3} STLS and RPA results for the screened ion potential at different values of the degeneracy parameter for $r_s=1.5$. The arrow indicates the change of the degeneracy parameter from 0.01 to 4. On this scale, the difference 
between the STLS and RPA results vanishes at large temperatures, i.e. at $\theta>1$.}
\end{minipage}
\label{fig:3}
\end{figure}

  \hspace{20mm}

\section{Discussion}\label{sec:4}

From the presented analysis of the screened ion potential at typical paramaters of non-ideal quantum plasmas we draw the following conclusions:\\

(i) Changes in the screening of the ion potential due to electronic correlations can have significant impact on the dynamics of the ionic subsystem if $\Gamma_i\gg1$. 
In experiments, a dense non-isothermal   plasma with a strongly coupled ionic subsystem can be realized. In this case, if $\Gamma_i\geq10$, the electronic non-ideality can have significant impact on the ion screening at $r_s<1$. Screening of the ion potential
at these densities was previously considered to be accurately described by the RPA. In order to confirm (or reject) the importance of the electronic non-ideality for the description of strongly coupled ions in quantum plasmas at $0.1<r_s<1$, 
an analysis of the dynamical and structural properties  of ions using both the ion potential in STLS approximation and RPA  (e.g. by molecular dynamics simulations) 
 should be performed.  \\
 
(ii) At $r_s\leq 2$, the ion potential in the STLS approximation has a shallow negative minimum which is missing in the calculations based on the QMC data for ground state. An accurate parametrization of the local field correction using 
 results of first principles QMC simulations  at finite temperature \cite{Tim_CPP, groth_prl} is required to extend the analysis of this feature to the warm dense matter regime. However, the aforementioned negative minimum due to the polarization of the electrons around the ion can be 
 fairly neglected at $r_s\leq1$ since then it becomes indistinguishable from the Friedel oscillations. 
 One can consider the negative minimum in the ion potential to be insignificant as long as the characteristic energy $E$ of the considered process in the plasma is $E\gg|\Phi_{\rm min}|$. 
 It is worth noting that strong attractions between like-charged particles exists in both the classical and quantum plasmas out of equilibrium, for example, the presence of a flow of mobile particles relative to the more inert species of particles \cite{Patrick_NJP, Moldabekov_PRE, Moldabekov_CPP, moldabekov_cpp_2016}. In Ref.~\cite{Shukla}, using quantum hydrodynamics (QHD), the attraction between like-charged ions in equilibrium plasmas due to the polarization of the surrounding electrons was reported. However, the accurate analysis on the basis of density functional theory \cite{Tim} and the density response function in the RPA \cite{POP2015} revealed that this attraction is an artifact arising from an incorrect treatment of the quantum non-loaclity in the QHD in the static case \cite{Michta}.\\
 
(iii) As expected, the analytical formulas for the ion potential derived within the RPA in the long wavelength approximation
are not able to correctly describe screening effects in non-ideal plasmas. However, the SM type potential can be improved to take into account correlation effects by considering, for example, an expansion of the local field correction \cite{Dandrea}
\begin{equation}\label{LFC}
 G(k)=\gamma k^2+\delta k^4,~\quad \gamma =-\frac{k_F^2}{4\pi e}\frac{\partial^2 n_ef_{\rm xc}}{\partial n_e^{\,2}}, ~\quad \delta=-\frac{\gamma^2}{2(1-g(0))},
\end{equation}
where $f_{\rm xc}$ is the exchange-correlation free energy per electron of the uniform electron gas, and $g(0)$ is the value of the electron-electron pair correlation function at $r=0$ \cite{Sjostrom}.  

Such consideration will lead to modifications of the coefficients in the SM potential, as it is discussed in Ref. \cite{Murillo} for the case $\delta=0$, $\gamma \neq 0$. 
A comprehensive and consistent analysis of this approach requires accurate QMC data for $f_{\rm xc}$, $\delta$ and a corresponding parametrization of the local field correction $G(k)$ over a wide range of wavenumbers $k$. Regarding the exchange-correlation free energy a highly accurate parametrization over the entire warm dense matter regime has been provided recently by Groth et al.~\cite{groth_prl}. In addition, first ab initio calculations of the local field correction at finite temperature have been successfully performed~\cite{dornheim4, groth}. Finally, an extension of the STLS for the description of dynamic (time dependent) correlations of classical systems was discussed by K\"ahlert et al. \cite{Hanno_1, Hanno_2} and a similar approach could be used for quantum systems.

Summarizing, the presented work highlights the importance of electronic non-ideality effects on the ion charge screening at $r_s\leq2$ as well as a high demand for accurate data on the static local field correction $G$ and its parametrization for dense plasmas at $\theta\precsim 1$.

\subsection*{Acknowledgments}
Zh. Moldabekov gratefully acknowledges funding from the German Academic Exchange Service (DAAD). This work has been supported by the Deutsche Forschungsgemeinschaft via grant SFB-TR24 project A9 and BO1366-10, and the Ministry of Education and Science of Kazakhstan under Grant No. 3086/GF4 (IPC-1) 2017.

%
%

\end{document}